\documentclass[runningheads]{llncs}

\usepackage{amsthm}
\theoremstyle{definition}
\usepackage{graphicx}
\usepackage[hidelinks]{hyperref}
\usepackage{amsfonts} 
\usepackage{graphicx}
\usepackage{floatrow}
\usepackage{amsmath}
\usepackage{algorithm}
\usepackage{algpseudocode}
\newfloatcommand{capbtabbox}{table}[][\FBwidth]
 \usepackage{multirow}

\DeclareMathAlphabet\mathbfcal{OMS}{cmsy}{b}{n}

\def\orcidID#1{\unskip$^{[#1]}$}
\begin{document}
\title{{One-Phase Batch Update on \\ Sparse Merkle Trees for Rollups}}
\author{Boqian Ma\orcidID{0009-0007-5247-3001}\and
Vir Nath Pathak\orcidID{0009-0008-4597-514X} \and
Lanping Liu\and
Sushmita Ruj\orcidID{0000-0002-8698-6709} 
}
\authorrunning{Ma et al.}
\institute{School of Computer Science and Engineering,\\University of New South Wales, Kensington, NSW 2052, Australia\\
\email{\{boqian.ma, vir.pathak, sushmita.ruj\}@unsw.edu.au} \\
\email{lanping.liu@unswalumni.com}}
\maketitle           
\begin{abstract}
A sparse Merkle tree is a Merkle tree with fixed height and indexed leaves given by a map from indices to leaf values. It allows for both efficient membership and non-membership proofs. It has been widely used as an authenticated data structure in various applications, such as layer-2 rollups for blockchains. zkSync Lite, a popular Ethereum layer-2 rollup solution, uses a sparse Merkle tree to represent the state of the layer-2 blockchain. The account information is recorded in the leaves of the tree. In this paper, we study the sparse Merkle tree algorithms presented in zkSync Lite, and propose an efficient batch update algorithm to calculate a new root hash given a list of account (leaf) operations. Using the construction in zkSync Lite as a benchmark, our algorithm 1) improves the account update time from $\mathcal{O}(\log n)$ to $\mathcal{O}(1)$ and 2) reduces the batch update cost by half using a one-pass traversal. Empirical analysis of real-world block data shows that our algorithm outperforms the benchmark by at most 14\%.

\end{abstract}

\section{Introduction}

Recent advances in distributed ledger technology have introduced a new paradigm of applications called ``decentralisation applications" (DApps) with new use cases in areas such as finance~\cite{kumar2020decentralising}~\cite{adams2021uniswap}, logistics~\cite{tijan2019blockchain}, and Internet-of-Things~\cite{panarello2018blockchain}. However, the increasing number of users and transactions on DApps has also exposed the key limitation of the scalability of their underlying public blockchain infrastructures~\cite{khan2021systematic}. Two of the largest public blockchians by market capitalisation\footnote{\url{https://coinmarketcap.com/} accessed on 23rd of August 2023.}, Bitcoin~\cite{nakamoto2008bitcoin} and Ethereum~\cite{wood2014ethereum}, can only process 7 and 29 transactions per second (TPS), which is far from their centralised payment provider counterpart, Visa, which claims to have the capacity to process 65,000 TPS~\cite{visa}. 

There are many ways to improve blockchain scalability. They can be broadly grouped into two categories: on-chain and off-chain. On-chain research involves changing the underlying blockchain infrastructure to achieve better scalability. Examples of on-chain research efforts include developing efficient consensus algorithms~\cite{kiayias2017ouroboros}~\cite{rocket2019scalable}, sharding~\cite{luu2016secure}~\cite{zamani2018rapidchain}, and changing block configurations~\cite{buterin2021_scaling_limit}. On the other hand, off-chain research efforts involve changing how we interact with the blockchain (L1). Instead of performing all activities on-chain, we offload the computation- and storage-intensive activities off-chain. Some existing solutions include State Channels~\cite{negka2021blockchain} , Plasma~\cite{poon2017plasma}, and rollups~\cite{Buterin2021_rollup_guide}. These scaling solutions are known as ``Layer-2" (L2) solutions.

The recent developments of L2 rollups such as zkSync Lite~\cite{zkSync_lite}, Aztec Network~\cite{aztec}, Loopering~\cite{loopering}, and Immutable X~\cite{immutable-x} has shown prominent results toward increasing transaction throughput on Ethereum. Rollups execute transactions off-chain and bundle the results of many L2 transactions into one L1 transaction. L1 cannot interpret L2 data, it only acts as a \textit{data availability layer} for L2 activity. Such techniques provide a reduction in computation to L1, while also massively decreasing the transaction fees as one L1 transaction fee is shared amongst all transactions bundled within it.

zkSync Lite~\cite{zkSync_lite}, a widely used and well-documented zero-knowledge rollup technique, has achieved a maximum observed TPS of 110~\cite{zksync-tps}, making it almost 6 times faster than Ethereum. Following the success of rollups, Ethereum has introduced a rollup-centric roadmap~\cite{Buterin2020_rollup_centric} specifically directing future scaling efforts on Ethereum to maximise the use of L2 rollups.

In an L2 rollup, there are generally \textit{operators} keeping the L2 state, processing L2 transactions and communicating with L1 through a smart contract.  \textit{Users} have \textit{accounts} and \textit{balances} of tokens. L2 users submit signed transactions to the operators, who then collect those transactions and form L2 blocks. 

Sparse Merkle trees (SMT) are widely used as authenticated data structures to keep state information in rollups because of their simplicity and effectiveness. The leaves of SMTs represent account-related information, such as balances and nonce. The root hash of SMTs is a succinct representation of the state of all account balances. Given a block of L2 transactions, the operators will calculate a new root hash based on the result of these transactions. Generally, the process of finding the root hash involves two parts: first, the account leaves need to be updated. Then, the new root hash is calculated by updating the paths from the updated leaves to the root. 

The current implementation of this in zkSync Lite is to first go through the transactions in a block sequentially to update the leaves individually and then calculate the root hash. This solution involves traversing the SMTs twice for every updated leaf, which is inefficient. We denote this as a two-phase algorithm.

To build on the above solution, this paper introduces the notion of\\ \texttt{BatchUpdate} on SMTs. The action of \textit{batching} is defined as processing transactions in a block all at once instead of individually. All accounts involved in transactions in a block are updated together in a batch.  Instead of traversing the SMTs twice, we propose a new algorithm to update the leaves and intermediate hashes at the same time by traversing the SMTs only once. We name this approach the one-phase batch update (OBU).
\subsubsection{Our Contributions.}
\begin{enumerate}
    \item We introduce an efficient SMT leaf update algorithm, \texttt{SMT.UpdateLeaf}, that improves account update time from $\mathcal{O}(\log n)$ to $\mathcal{O}(1)$.

    \item Building on this, an SMT batch update algorithm, \texttt{SMT.BatchUpate}, is proposed to calculate the root hash of an SMT, reducing the total number of traversals by 50\% from $\mathcal{O}(k\log n) + \mathcal{O}(k\log n) H$ to $\mathcal{O}(k\log n)H$, where $k$ is the number of updates in a batch, $n$ is the total number of leaves in the SMT, and $H$ is a hash operation.\footnote{Code at: \url{https://github.com/Boqian-Ma/one-phase-batch-update-SMT}}

    \item Performance analysis of our proposed algorithm was conducted using both micro- and macro-benchmarks in single and multi-threaded scenarios. 

    \item In real-world macro-benchmark data, our algorithm outperformed the benchmark by up to 14\%. 
\end{enumerate}

\subsubsection{Organisation.}
The rest of the paper is organised as follows. Section \ref{sec:preliminary} introduces the preliminary information. Next, Section \ref{sec:related_work} discusses some related work. In Section \ref{sec:obu}, we introduce the batch update algorithm. Section \ref{sec:evaluation} outlines our experimental results, followed by the conclusion and discussion in Section \ref{sec:conclusion}. 

\section{Preliminaries}\label{sec:preliminary}

\subsection{Leaf Operation}\label{sec:leaf_operations} 
\begin{definition}[Leaf Operation]
    Given a Merkle tree (MT), $T$, with $n$ leaf nodes $L = \{\mathrm{leaf}_0, \cdots, \mathrm{leaf}_{n-1}\}$ and their corresponding data items\\$D = \{d_0, \cdots, d_{n-1}\}$ where $\mathrm{leaf}_j=H(d_j)$, a leaf operation\\$o^{j} \in \{\texttt{InsertLeaf, UpdateLeaf, RemoveLeaf}\}$ where $0 \leq j < \vert D \vert$, is a function that modifies the value of $\mathrm{leaf}_j$. \texttt{InsertLeaf} inserts a new leaf, given by $\mathrm{leaf}_j=H(d_j)$, into the tree, \texttt{UpdateLeaf} updates the value of $\mathrm{leaf}_j$, and \texttt{RemoveLeaf} removes $\mathrm{leaf}_j$ and $d_j$ from the tree and $D$ respectively.

\end{definition}

\subsection{Sparse Merkle Tree}
\begin{definition}[Sparse Merkle Tree]
An SMT is an MT with a fixed depth of $N$, and indexed leaves. Data items $D= \{d_0, \cdots, d_{n-1}\}$ where $n < 2^{N}$ are stored in a map, $M: \{0,1\}^{2^{N}} \rightarrow D$ mapping from leaf indices to data items. An SMT is defined by the following set of algorithms on $M$:
\begin{enumerate}
    \item $\texttt{Gen($N$)} \rightarrow \texttt{SMT}$: Algorithm that generates an empty SMT given a depth $N$.
    \item  $\texttt{SMT.Commit($M$)} \rightarrow \texttt{R}'.$ Deterministic algorithm that inserts every key-value pair in $M$ into the tree and returns the new root hash.  
    \item  $\texttt{SMT.ApplyOp($o^i$)} \rightarrow \texttt{R}'$ Deterministic algorithm that applies the leaf operation $o^i$ and returns a new root hash $R'$. \texttt{SMT.ApplyOp($o^i$)} can be further categorised into three methods depending on the operation type. They are \texttt{SMT.InsertLeaf($o^i$)}, \texttt{SMT.UpdateLeaf($o^i$)}, and \texttt{SMT.RemoveLeaf($o^i$)}. A description of each of these operations can be found in Section \ref{sec:leaf_operations}.
    \item $\texttt{SMT.MemberWitnessCreate($i$)} \rightarrow w_i$: Deterministic algorithm that returns the Merkle proof of $M(i)$ consisting of a list of siblings nodes from $\mathrm{leaf}_i$ to the root.
    \item $\texttt{SMT.MemberVerify($w_i$, $d_i$)} \rightarrow \{true, false\}$: Deterministic algorithm that verifies whether $d_i$ is a member of $M$.
\end{enumerate}
\end{definition}

SMTs have the same membership-proof construction as regular Merkle trees. However, proving non-membership is more efficient on SMTs than on Merkle trees, since a non-membership for a key $k$ in an SMT is the membership proof of the default value.

\begin{figure}[ht]
\centering
\includegraphics[width=0.8\textwidth]{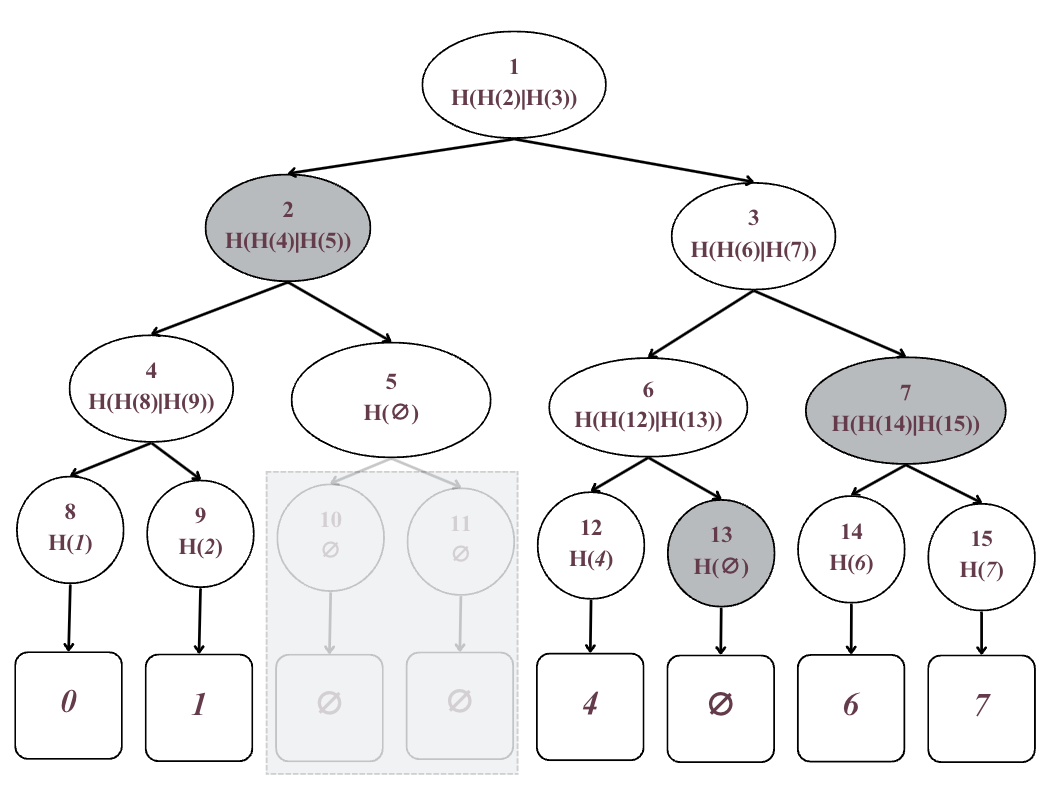}
\caption{A SMT of 3 levels. The ovals represent internal nodes. The squares represents its value mapping $M$, where the numbers are the keys of $M$ and the leaf indices. The default value is represented as $\emptyset$. The highlighted nodes form $\mathrm{leaf}_4$'s membership proof. Since $\mathrm{leaf}_2$ and $\mathrm{leaf}_3$ are empty, everything below their highest common parent, $\mathrm{node}_5$, are pruned to increase storage efficiency.}
\label{fig:smt}
\end{figure}

\subsubsection{Space optimisation.} Instead of storing the full SMT of $2^{N+1}-1$ nodes, Bauer~\cite{bauer2004proofs} presents a memory efficient way of storing an SMT by pruning empty sub-trees. Referring to node 5, Figure \ref{fig:smt}, following Bauer's proposal, the subtree of node 5 is replaced with the default hash. As such, the space can be greatly reduced. 

\section{Related Work}\label{sec:related_work}
This section introduces zkSync Lite~\cite{zkSync_lite} and its relevant SMT root hash update algorithm, which we use as our benchmark. 

\subsubsection{zkSync Lite}
zkSync Lite~\cite{zkSync_lite} is an L2 rollup solution developed by Matter Labs~\cite{matter_labs}. It supports simple transaction types including transfer or swap of ERC-20~\cite{erc20} tokens, and ERC-721~\cite{erc721} token minting. Like most L2 solutions, zkSync Lite has two main components: on-chain and off-chain. The on-chain component includes several Solidity Smart Contracts deployed \footnote{\url{https://etherscan.io/address/0xaBEA9132b05A70803a4E85094fD0e1800777fBEF}} on Ethereum L1. The off-chain component includes several micro-services that facilitate L2 transaction executions and SNARK~\cite{gabizon2019plonk} generation. Detailed descriptions of the zkSync Lite design are given in the Appendix \ref{appx:zksync_lite_design}
\subsubsection{Account Tree Construction.}
SMTs are used in three places in zkSync Lite: account tree \footnote{ \url{https://github.com/matter-labs/zksync/blob/master/core/lib/types/src/lib.rs\#L84}}
, circuit account tree, and balance tree. The account tree is the main data structure that keeps track of the account balances of its users. The circuit account tree and the balance tree are derived from the account tree and are used to build zero-knowledge block proofs. Here, we give descriptions of the account tree in zkSync Lite. 

The account tree is an SMT of depth $N=24$. As such, it can store up to $2^{25}-1$ accounts. The accounts are stored in a map $M$, mapping from leaf indices to accounts. Each internal node, $\mathrm{node}_j$, where $1 \leq j \leq 2^{N+1}$, $\mathrm{node}_j$'s direct children are $\mathrm{node}_j$'s children are given by $\mathrm{node}_{2j}$ and $\mathrm{node}_{2j+1}$ and $\mathrm{node}_j = H(\mathrm{node}_{2j} \| \mathrm{node}_{2j+1})$. $\mathrm{node}_j$ is also known as $\mathrm{node}_{2j}$ and $\mathrm{node}_{2j+1}$'s parent node. The root of the tree is $\mathrm{node}_1$, which also corresponds to the digest of $T$. 

Each leaf node $\mathrm{leaf}_k$ where $0 \leq k < 2^{N}$ , corresponds to a key $k$ and is labelled with the value associated with that key if it exists or the hash of a default value otherwise. Formally, if $v = M(k)$ exists, $\mathrm{leaf}_k = v$, else $\mathrm{leaf}_k = \textit{default}$, where \textit{default} is a predefined default value. 

On the $N^{\mathrm{th}}$ level of $SMT$ (i.e. the leaf level), given by the set of $2^N$ nodes $\{\mathrm{node}_q\}^{2^N}$ where $2^N \leq q < 2^{N+1}$, each $\mathrm{node}_j$ corresponds to a key $k = (1 << N) + q$ and is labelled with the hash of the value associated with that key if it exists, or the hash of a default value otherwise. Formally, if $v = M(k)$ exists, $\mathrm{node}_q = H(v)$, otherwise $\mathrm{node}_q = H(\textit{default})$. 

For simplicity, we denote the nodes at the leaf level by $L = \{\mathrm{leaf}_0, \cdots, \mathrm{leaf}_k\}$ where $0 \leq k < 2^{N}$.

\subsubsection{Root Hash Update Algorithm.}
Here we outline the root hash update algorithm implemented in zkSync Lite given a list of leaf operations. This algorithm is divided into two phases. Consider an account tree $T$ and a list of $k$ operations $O = \{o^{j}\}^k_{j\in[0, 2^{N})}$. The first phase updates the leaves to their new values. For each operation $o^{j} \in O$, the algorithm traverses $T$ from the root to $\mathrm{leaf}_j$ and performs the operation. For example, if $o^j$ was an \texttt{update} balance operation, then the balance of $\mathrm{leaf}_j$ is updated accordingly. At the end of this phase, all accounts affected by $O$ are updated. Note that when a leaf is updated to a new value, all nodes in its parent path need to be recomputed. This phase does not concern the hash calculation and takes $\mathcal{O}(k\log n)$ running time to perform $k$ updates.

The second phase re-computes the hashes of affected paths and returns the new root hash. To compute the root hash, the algorithm traverses left and right recursively from $T$'s root to retrieve or compute the child hashes. Recursion terminates when 1) an updated leaf is reached or 2) when all the child leaves of the current nodes are unchanged from the first phase. In case 1), the leaf hash is calculated and returned. In case 2), the current node hash is returned. As a result of this recursive algorithm, the new root hash is calculated. This phase takes $\mathcal{O}(k\log n)H$ running time, where $H$ is the running time of the chosen hash function. Together, the root hash calculation process takes $\mathcal{O}(k\log n) + \mathcal{O}(k\log n)H$.

The first phase occurs in the block producer module, while the second phase occurs in the root hash calculator module. In the actual implementation, these two phases are completed in two separate micro-services. The first phase occurs in the ``block producer" module, where the leaves are updates. Then, the second phase happens in the ``root hash calculator" module, where the new root hash is computed. This separation takes the hash calculation computation overhead away from the main service.

\subsubsection{Inefficiencies.} \label{sec:inefficiency}
Above we described a two-phase algorithm implemented in zkSync Lite to update the root state of the account tree given a list of $k$ leaf operations. As stated in the zkSync Lite code base \footnote{\url{https://github.com/matter-labs/zksync/blob/master/core/bin/zksync_core/src/state_keeper/root_hash_calculator/mod.rs\#L21}}, there exists a bottleneck that constrains the speed of the block producer producing blocks. If the block producer's speed exceeds the speed of root hash calculation, then the job queue for the root hash calculator will increase indefinitely. Furthermore, we observe that for each operation $o^{j}\in O$, the path between the updated $\mathrm{leaf}_j$ and root is traversed twice. The first traversal occurs when updating the account values and the second time occurs when calculating the root hash. 

\section{One-Phase Batch Update on Sparse Merkle Tree}\label{sec:obu}
In this section, we first outline the basic functionalities of the three leaf operations, \texttt{SMT.InsertLeaf, SMT.UpdateLeaf} and \texttt{SMT.RemoveLeaf}. Then, we introduce a more efficient algorithm, \texttt{SMT.BatchUpdate($\mathbf{O}$)$\rightarrow R'$} that takes in a list of operations and returns the new SMT root. The pseudocode is outlined in Algorithm \ref{alg:cap}. 

\subsection{Leaf Operation Algorithms}
\begin{enumerate}
    \item \texttt{SMT.InsertLeaf($\mathrm{leaf}_j$)} is a deterministic algorithm that inserts $\mathrm{leaf}_j$ into the SMT by traversing from the root. It has a runtime of $\mathcal{O}(\log n)$.
    \item \texttt{SMT.UpdateLeaf($\mathrm{leaf'}_j$)} is a deterministic algorithm that updates the value of $\mathrm{leaf}_j$ to $\mathrm{leaf'}_j$. This algorithm assumes the existence of $v = M(j)$. As such, we can complete this algorithm in $\mathcal{O}(1)$.
    \item \texttt{SMT.RemoveLeaf($j$)} is a deterministic algorithm that updates the value of $\mathrm{leaf}_j$ to \textit{default}. Similar to \texttt{SMT.UpdateLeaf}, it assumes the existence of $v = M(j)$ and can be completed in $\mathcal{O}(1)$.
\end{enumerate}

\subsection{Batch Update Algorithm}
\texttt{SMT.BatchUpdate} is based on bottom-up binary tree level-order traversal using a queue data structure. It is broken down into two parts. In the first part (lines 4-9), we update the leaf nodes. In the second part (lines 10-19), we re-calculate the hashes of nodes in the affected paths in a bottom-up fashion and eventually return the new root hash. \textbf{T.cache} is a list of nodes that make up the tree. 

Referring to lines 4 to 9, we first initialise an empty set $parent\_set$, which we will use to store the indices of the direct parent nodes of the leaves that we updated. We use a set data structure to avoid duplicated parents (i.e. if we update both $\mathrm{node}_4$ and $\mathrm{node}_5$, then the parent node of both nodes, $\mathrm{node}_2$, will only be added to the parent set once). Next, for each $o^{j} \in O$, we apply $o^{j}$ to the value $M(j)$, calculate the new hash of $\mathrm{leaf}_j= H(M(j))$ and add $\mathrm{leaf}_j$'s parent node's index to $parent\_set$. As a result of performing all operations, $parent\_set$ is filled with a set of node indices at a level above the leaf level (i.e. $N-1$).

Referring to lines 10-19, given $\mathrm{parent\_set}$, we first empty them into a queue $\mathrm{current\_level}$, which represents the indexes of the nodes we are updating. Next, for each $i \in \mathrm{current\_level}$ we calculate and update $H(\mathrm{node}_i)$ by retrieving $i$'s children hashes $H(\mathrm{node}_{2i})$ and $H(\mathrm{node}_{2i+1})$ from $T$. We are guaranteed to retrieve the most recently updated children's hashes because when we process indexes at level $n$ where $0 \leq n\le N$, nodes in $n+1$ have already been updated. Then, we add $\mathrm{node}_i$'s parent index $\mathrm{node}_{\lfloor i/2 \rfloor}$ to $parent\_set$. We repeat this process until we reach the root level of $T$. As a result, $\mathrm{node}_1$ (i.e. the root) will be updated and returned.

\begin{algorithm}[ht]
\caption{Sparse Merkle Tree Batch Update}\label{alg:cap}
\begin{algorithmic}[1]
\State \textbf{Input:} Sparse Merkle Tree \textbf{T} of depth $N$, List of leaf operations $\textbf{O} = \{o^j\}^{k}_{j\in[0,2^N)}$ of size $k$.
\State \textbf{Output:} Root Hash \textbf{H}
\Procedure{SMTBatchUpdate}{$\textbf{T}, \textbf{O}$}
\State parent\_set $\gets$ Set() 
\ForAll{${o}^j \in \mathbf{O}$}
    \State perform operation ${o}^j$ on $\mathrm{leaf}_j$
    \State calculate the new hash of $\mathrm{leaf}_j$ and update the value in $\mathbf{T}$
    \State parent\_set.add($\mathrm{leaf}_j$.parent)
\EndFor
\While{parent\_set is non-empty} 
    \State current\_level = empty(parent\_set)
    \For{parent $p_i$ in current\_level} 
        \State left\_child\_hash = get\_child\_hash($p_i$.left)
        \State right\_child\_hash = get\_child\_hash($p_i$.right)
        \State calculate the new hash of $p_i$ by using left\_child\_hash and right\_child\_hash $p_i$ and update the value in T.
        \State parent\_set.add($p_i$.parent)
    \EndFor
\EndWhile

\Return T.cache[ROOT\_index]
\EndProcedure
\end{algorithmic}
\end{algorithm}

\subsubsection{Example.}\label{sec:example}
\begin{figure}[ht]
\centering
\includegraphics[width=1\textwidth]{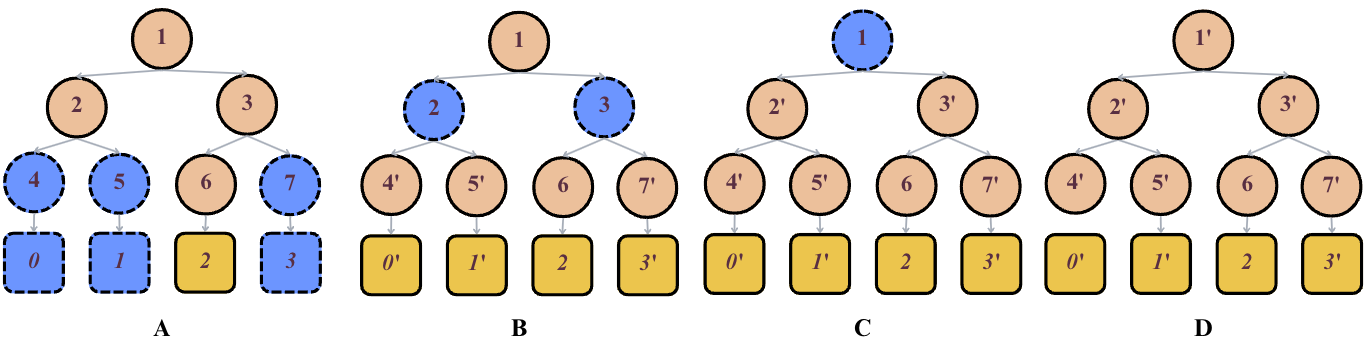}
\caption{An illustration of the one-phase batch update example provided in section \ref{sec:example}. Circle nodes are internal nodes, square nodes are data items with leaf indices, dotted boarders represent the nodes that are currently in the queue, and an apostrophe on a number represents the updated state of a node. }
\label{fig:one-phase-example}
\end{figure}

To illustrate the above algorithm, consider an SMT of depth 2 and a list of operations $O=\{o^{0}, o^{3}, o^{1}\}$. Figure \ref{fig:one-phase-example} (A) shows the leaf level nodes that are affected by $O$, they are $L_2 = \{\mathrm{node}_4, \mathrm{node}_5, \mathrm{node}_7\}$ and their corresponding values in $M$ (i.e. $M(0), M(1), M(3)$). 

As a result of updating $M$ and re-hashing $L_2$ nodes, \ref{fig:one-phase-example}.B shows the updated leaf nodes and $M$, and the parent nodes of $L_2$ which are $L_1 = \{\mathrm{node}_2, \mathrm{node}_3\}$ as dotted borders. Now, to re-hash $\mathrm{node}_2$, we retrieve $\mathrm{node}_2$'s children nodes which are $\mathrm{node}_4'$ and $\mathrm{node}_5'$. The same can be done for $\mathrm{node}_3$. Figure \ref{fig:one-phase-example} (C) shows the result of re-hashing $L_1$, and the parent nodes of $L_1$, which is $\mathrm{node}_1$. In the end, Figure \ref{fig:one-phase-example} (D) shows the final result of the algorithm and a new root hash.

\subsection{Comparison}
Table \ref{tab:obu_complexity_comparision} compares the performance of the baseline and OBU for different types of leaf operations, \texttt{SMT.Commit}, and \texttt{SMT.BatchUpdate}. The table assumes an SMT of $n$ leaves and a list of $k$ operations. Although our \texttt{SMT.BatchUpdate} has the same asymptotic time complexity, it is more efficient because the improvement in \texttt{SMT.UpdateLeaf} and \texttt{SMT.RemoveLeaf}. Furthermore, the space complexity of our algorithm remained the same as the baseline algorithm, which is $\mathcal{O}(2^{N})$.

\begin{table}[ht]
    \centering
    \begin{tabular}{|c||c|c|}
        \hline
        Method & zkSync Lite~\cite{zkSync_lite} & OBU\\
        \hline
        \hline
        \texttt{SMT.InsertLeaf} & $\mathcal{O}(\log n)$ &  $\mathcal{O}(\log n)$\\
        \hline
         \texttt{SMT.UpdateLeaf} & $\mathcal{O}(\log n)$  & $\mathbfcal{O}$\textbf{(1)}\\
         \hline
        \texttt{SMT.RemoveLeaf} & $\mathcal{O}(\log n)$ &  $\mathbfcal{O}$\textbf{(1)}\\
        \hline
        $\vert w_i \vert$ & $\mathcal{O}(\log n)$ & $\mathcal{O}(\log n)$\\
        \hline
        \texttt{SMT.Commit} & $\mathcal{O}(k\log n)H$ &  $\mathcal{O}(k\log n)H$ \\
        \hline
         \texttt{SMT.BatchUpdate} & $\mathcal{O}(k\log n) + \mathcal{O}(k\log n)H$ &  $\mathcal{O}(k\log n)H$ \\
        \hline
    \end{tabular}
    \caption{Asymptotic complexity comparison between OBU and the baseline. $n$ is the number of leaves, $k$ is the number of operations in a block, and $H$ is a hash operation.}
    \label{tab:obu_complexity_comparision}
\end{table}

\section{Experimental Analysis}
We performed both micro- and macro-benchmarks to compare our algorithm with the benchmark. The micro-benchmarks consisted of simple leaf operations in single-threaded and multi-threaded settings. The macro-benchmark compared the performance of the algorithms on real-world block data from zkSync Lite. This section describes the experimental setup, the dataset used for the macro-benchmark, and the multi-threading optimisation for \texttt{SMT.BatchUpdate}.

\subsection{Experimental setup}
zkSync Lite is implemented in the Rust programming language~\cite{klabnik2023rust} as an open source project on Github \footnote{https://github.com/matter-labs/zksync}. We implemented Algorithm \ref{alg:cap} on top of the existing repository. Further, we also optimised our implementation for multi-threading computation using the Rayon~\cite{rayon} library in rust. 

The experiments are performed on an \textit{AWS c5.12xlarge Debian, 48 CPU, 96 GiB memory} virtual machine. The SMT we used for our experiments has a depth of 24, which is the same depth as the one in zkSync Lite. For each experiment, we performed 10 runs and reported the average run time. The main metric we use to compare performance is the percentage decrease in run time given by 
$$
\% \text{decrease in running-time} = \frac{\text{new running-time} - \text{old running-time}}{\text{old running-time}}.
$$

\subsection{Dataset Collection}
The macro-benchmark dataset contains 100 (block \#299246- \#299346) recent blocks and their transactions which are collected through the zkSync Lite API~\cite{zksync_api_doc} and the zkSync Lite block explorer~\cite{zksync_explorer}. 

Of the 8376 transactions collected, 3971 are swap transactions, 1897 are transfer transactions, 1428 are MintNFT transactions, 766 are ChangePubKey transactions, 266 are deposit transactions, 47 are withdraw transactions, and only 1 is a WithdrawNFT transaction. Details of these transaction types can be found in Appendix \ref{appx:zksync_tx_types}.

More than 70\% of the transactions are dominated by ERC-20 token transactions. To keep the experiments simple, we only considered the Transfer and Swap transaction. 

We also noticed that the transaction count for each block is inconsistent. The maximum number of transactions observed was 133 while the minimum was 74. This is the result of a combination of the gas limit reached and the appearance of \textit{Priority Transactions} such as Deposit and Withdraw during transaction processing, which will cause the current block to be sealed and committed as soon as it is processed. 

We observed that there are many highly active accounts. In block \# 299273, out of 92 transactions, one leaf was included in 48 transactions, taking up more than half of the block space. On average, each account produced 2.5 transactions in our dataset.

\begin{figure}[ht]
\begin{floatrow}
\ffigbox{%
  \includegraphics[width=0.5\textwidth]{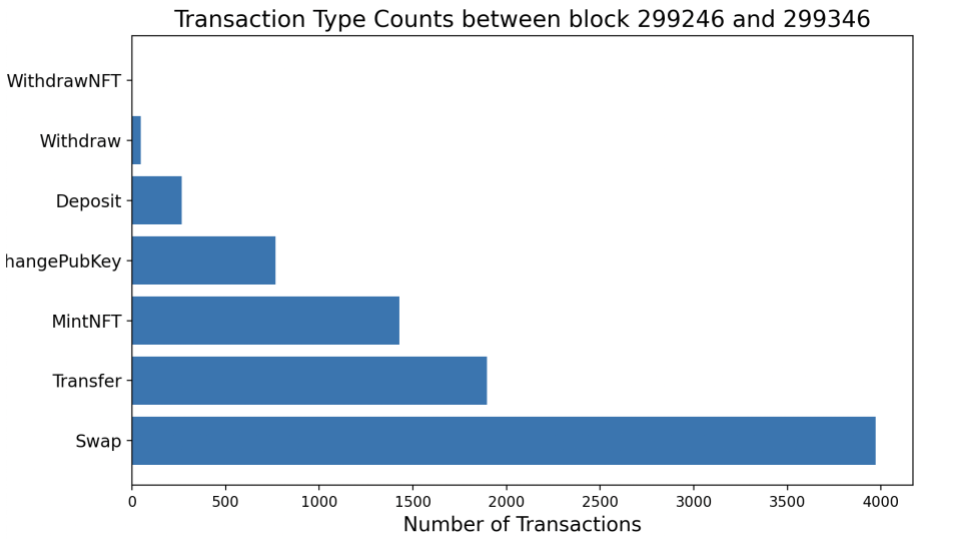}
}{%
  \caption{Transaction count by type between blocks 299264 and 299364}%
  \label{fig:tx_distribution}
}
\capbtabbox{%
    \small
  \begin{tabular}{|l|r|}
    \hline \textbf{Statistic} & \textbf{Value}\\
    \hline
    \hline Total \# txs & 8376 \\
    \hline Max tx count in a block & 133 \\
    \hline min tx count in a block & 74 \\
    \hline Average tx count in a block & 83 \\
    \hline Unique accounts & 3322 \\
    \hline Average tx per account & 2.5 \\
    \hline
    \end{tabular}
    \label{table:dataset_info}
}{%
  \caption{Macro-benchmark dataset information of zkSync Lite blocks 299246 - 299346}
}
\end{floatrow}
\end{figure}

\subsection{Multi-threading Optimisation}
Both the baseline and OBU can be optimised for multi-threading. In the baseline, threads can be created in the recursive stage by visiting the child nodes. In OBU, a thread can be created for every node that requires re-hashing in a level. Note that in the baseline, the threads are nested as the tree is traversed deeper, whereas in OBU, there are no nested threads. 

\section{Evaluation}\label{sec:evaluation}
\subsection{Micro-benchmarks}
In Section \ref{sec:leaf_operations} we gave three categories of leaf operations: \texttt{SMT.UpdateLeaf},\\ \texttt{SMT.InsertLeaf}, and \texttt{SMT.RemoveLeaf}. In the micro-benchmarks, we performed simple leaf operations to demonstrate the effectiveness of our proposed algorithm. Without losing generality, we did not include experiments for \texttt{SMT.RemoveLeaf} operation as the implementation is similar to \texttt{SMT.UpdateLeaf}.\\
\subsubsection{With multi-threading.}
\begin{figure}[ht]
\centering
\includegraphics[width=1\textwidth]{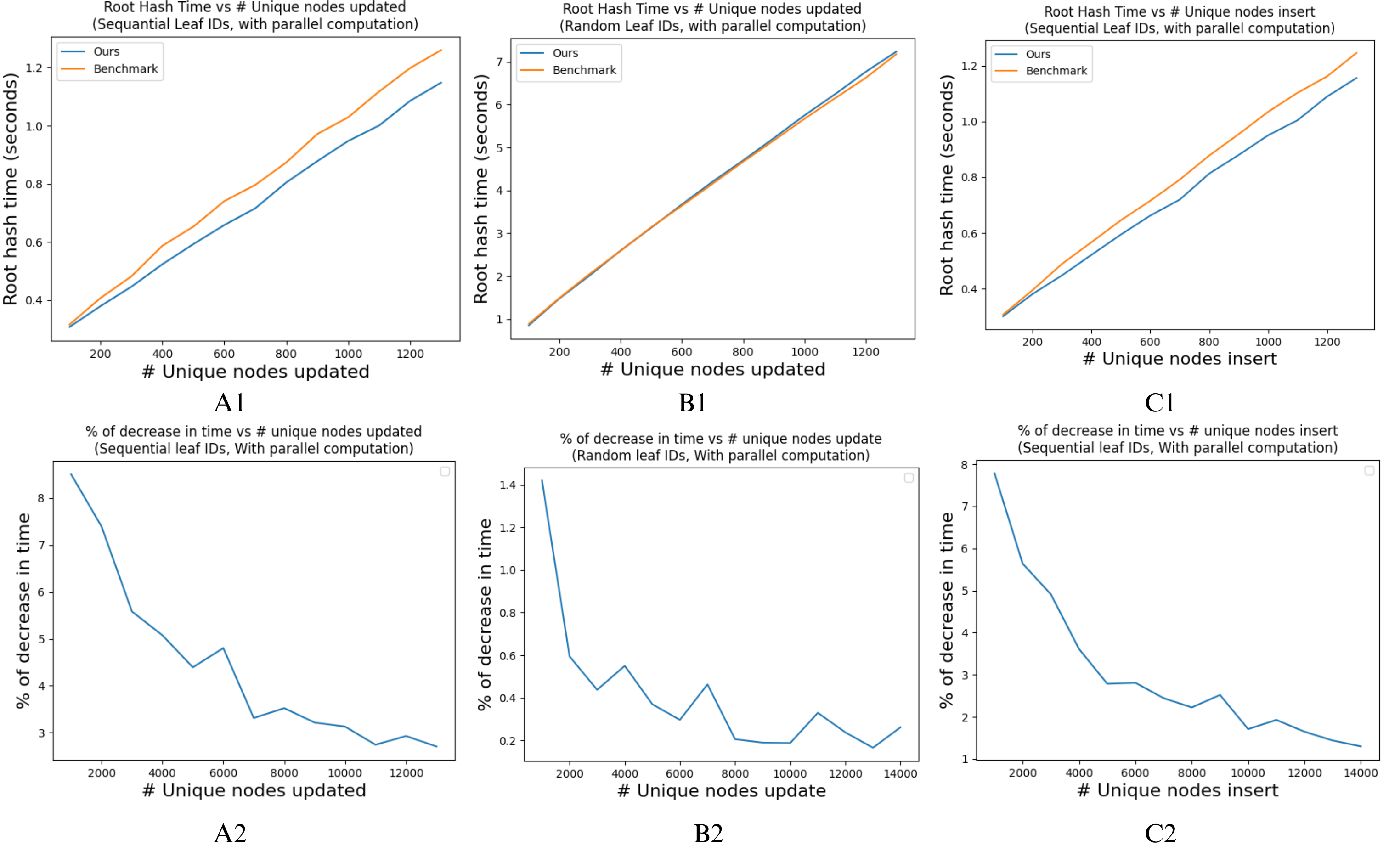}
\caption{Top row: root hash time in seconds comparison between benchmark and OBU with various operation types. Bottom row: percentage decrease in root hash time with various operation types. (with multi-threading)}
\label{fig:micro-multi-thread}
\end{figure}

Figure \ref{fig:micro-multi-thread} shows the performance comparison when multi-threading is enabled. In Figure \ref{fig:micro-multi-thread} (A1), when the update operations are applied to leaves with sequential IDs, we see that OBU outperforms the baseline. We also note that the gap in runtime is increasing by an increasing factor. This is expected because given $k$ update operations, the baseline spends $\mathcal{O}(k\log n)$ on traversal and update, while with OBU, the update time is linear with respect to $k$ (i.e. $\mathcal{O}(k)$ update time).

In Figure \ref{fig:micro-multi-thread} (A2), we see that when the number of operations is small, we see a larger percentage decrease in running time and as the number of operations increases (10\% decreases for 1000 updates), \% decrease in running time shows exponential decay. The initial large percentage decrease relates to how the two algorithms use multi-threading. In the baseline, threads can be nested as deep as 24 levels, which can cause high computation overhead, whereas in OBU, there is no such problem because threads end when the currently traversed level is finished. Furthermore, the diminishing trend in Figure \ref{fig:micro-multi-thread} (A2) can be explained by hardware limitations. In OBU, as the number of nodes we process on each level increases, the number of concurrent threads becomes insignificant compared to the number of nodes we need to process. 

Figure \ref{fig:micro-multi-thread} (B1) shows the runtime difference when the update operations are applied to random leaf IDs taken from a uniform distribution. We note that the improvement in runtime is worse visually compared to Figure \ref{fig:micro-multi-thread} (A1). This is because when leaf IDs are randomly assigned, there are fewer common parents. As such, the amount of computation of OBU approaches the baseline. However, we also note that the trend shown in Figure \ref{fig:micro-multi-thread} (B2) is consistent with Figure \ref{fig:micro-multi-thread} (A2) when it comes to the percentage of decrease in running time. 

Figure \ref{fig:micro-multi-thread} (C1) shows the runtime difference for when insert operations are applied to leaves with sequential IDs. Both Figure \ref{fig:micro-multi-thread} (C1) and Figure \ref{fig:micro-multi-thread} (C2) show consistent trends as Figure \ref{fig:micro-multi-thread} (A1) and Figure \ref{fig:micro-multi-thread} (A2) respectively.

\subsubsection{Without multi-threading.}
\begin{figure}[ht]
\centering
\includegraphics[width=1\textwidth]{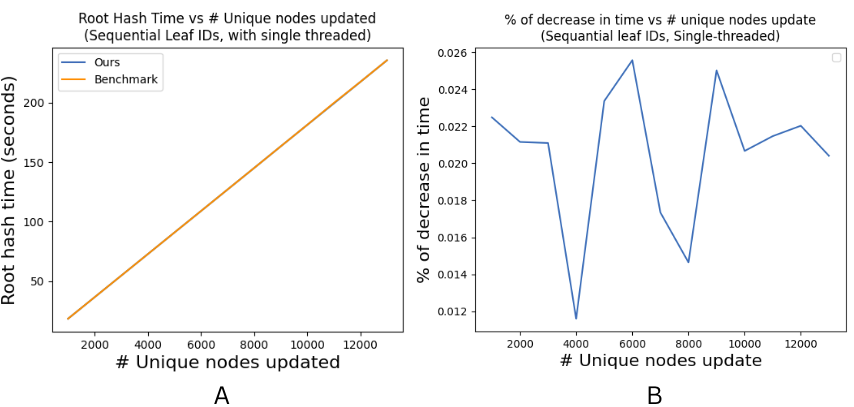}
\caption{A) Update operations on leaves with sequential leaf IDs performance comparison, no visible difference. B) Percentage decrease in running time with OBU compared with baseline. (Single-threaded)}
\label{fig:single-thread}
\end{figure}

Figure \ref{fig:single-thread} (A) shows the running time comparison between the benchmark and OBU when running on a single thread. We note that there is no visible performance improvement because the tree traversal time $\mathcal{O}(k\log n)$ is insignificant compared to the hashing time. This is further demonstrated in Figure \ref{fig:single-thread} (B) when we only observe a slight improvement in the percentage decrease in running time. 

\subsection{Macro-benchmark}
\begin{figure}[ht]
\centering
\includegraphics[width=1\textwidth]{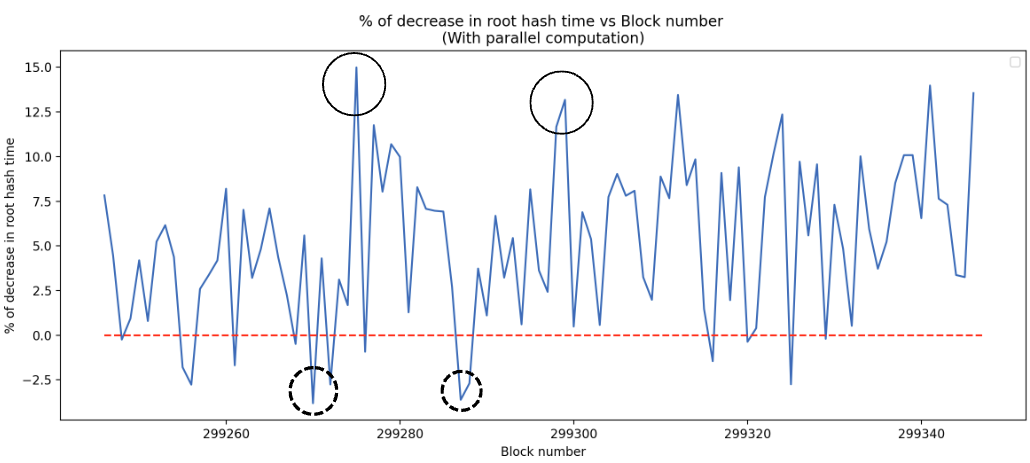}
\caption{Percentage decrease in running time on zkSync Lite block data. The dotted horizontal line is when the percentage of decrease is 0.}
\label{fig:macro}
\end{figure}

We macro-benchmark the performance of OBU with the baseline using zkSync Lite block data. As shown in Figure \ref{fig:macro}, OBU almost always outperforms the baseline. Overall, OBU performed, on average, 5.12\% faster than the baseline, with the highest percentage of decrease in time being 14\%. Next, we analysed the blocks that exhibited the highest/lowest performance improvement. Our observations are as follows:
\begin{enumerate}
    \item In the block with large percentage of decrease in time (i.e. in solid circles in Figure \ref{fig:macro}) we notice that most transactions in the block affected very few accounts. This corresponds to a faster running time because OBU does not repeatedly traverse the same account
    \item In the blocks with negative percentage of decrease in time (i.e. in dotted circles in Figure \ref{fig:macro}), updates are spread across multiple accounts instead of just a few accounts. 
\end{enumerate}

\begin{table}[ht]
\centering
\begin{tabular}{|c|c|}
\hline
\textbf{Statistic} & \textbf{Runtime Reduction (\%, ms)}\\
\hline
Mean & 5.12\%, 25.56\\
\hline
Medium & 5.24\%, 26.73\\
\hline
Standard Deviation & 4.39\%, 21.55\\
\hline
Variance & 19.24\%, 464.68\\
\hline
Minimum & -3.81\%, -20.67\\
\hline
Maximum & 14.99\%, 69.79\\
\hline
Range & 18.80\%, 90.46\\
\hline
\end{tabular}
\caption{Result statistics from the macro-benchmark.}
\label{table:macro_stats}
\end{table}

\section{Conclusion and Discussion}\label{sec:conclusion}
In this paper, we presented and evaluated OBU, a batch update algorithm on sparse Merkle trees. The improvement can be summarised as follows. OBU achieved a 50\% decrease in run time by traversing the tree once instead of twice. OBU uses threads more efficiently compared to the implementation presented in zkSync Lite. This could reduce the hardware requirement to run an L2 operator. More specifically, OBU reduced the run time by 50\% for the \texttt{SMT.InsertLeaf} operation. For \texttt{SMT.UpdateLeaf} and \texttt{SMT.RemoveLeaf} operations, the running time is reduced from $\mathcal{O}(\log n)$ to $\mathcal{O}(1)$. 
\subsubsection{High Frequency Transaction Applications.} 

The second improvement will directly benefit applications with a higher frequency of transactions. Suppose that a block has $k$ transactions that affect a single account. Instead of traversing the SMT $k$ times in $\mathcal{O}(k\log n)$ runtime, OBU will complete the operations in $\mathcal{O}(k)$ runtime. This is evident in block \#299275\footnote{\url{https://zkscan.io/explorer/blocks/299275}} where 29 of the 47 transfer/swap transactions in block. In this case, OBU achieved a 14.9\% decrease in running time.  

\subsection{Future Work} For future work, we first want to perform more integration tests in zkSync Lite to better understand the advantages and drawbacks of OBU. Next, we wish to see how our research can improve zkEVM, which is another prominent blockchain scaling direction. Then, we want to see how our research may be used for batch update in other authenticated data structures, such as Vector Commitment schemes~\cite{catalano2013vector}. 

\subsubsection{Acknowledgement.} The authors extend their thanks to Sean Morota Chu, Ziyu Liu, Nhi Nguyen, and Tim Yang for invaluable feedback on the manuscript, Barak Saini for helping us understand zkSync Lite, and Hao Ren for \LaTeX{} formatting advice.

\bibliographystyle{splncs04}
\bibliography{ref}
\appendix
\section{zkSync Lite Details}

\subsection{Design} \label{appx:zksync_lite_design}
Like most L2 solutions, zkSync Lite has two main components: on-chain and off-chain. The on-chain component includes several Solidity Smart Contracts deployed \footnote{https://etherscan.io/address/0xaBEA9132b05A70803a4E85094fD0e1800777fBEF} on the Ethereum mainnet. The off-chain component includes several microservices that facilitate L2 transaction executions and SNARK generation. 

\subsection{On-Chain}
The on-chain component has three main contracts. 

The first one is the \textsf{zkSync} main contract. It stores L1 user funds, bridges funds between L1 and L2 with \texttt{Deposit} and \texttt{Withdraw} transactions, accepts committed blocks and block proofs from the operator, verifies block proofs, and process withdrawal transactions by executing blocks. Users can deposit \$ETH or ERC-20 tokens. However, the allowed ERC-20 tokens are determined by the Security Council.

The second Smart Contract is \textsf{Verifier}. Given a committed block and a proof, the \textsf{Verifier} contract verifies the proof to determine the validity of the state transition caused by the transactions in the block. 

The third Smart Contract is \textsf{Governance}. It has the functionalities to add (but not remove) ERC-20 tokens to the whitelisted tokens, change the set of operators, and initiate the upgrade of the contracts. 

When L1 users wish to deposit/withdraw their funds to/from L2, they can interact directly with the \textsf{zkSync} main contract.

\subsection{Off-Chain}
The off-chain component is divided into two main sub-components. The server and the prover. An operator needs to run both sub-components in order to create L2 blocks.
\subsubsection{Server.}
The Server has the following modules~\cite{zkSync_lite}:
\begin{enumerate}
    \item Ethereum Watcher:  module to monitor on-chain operations.
    \item State Keeper: module to execute and seal blocks.
    \item Memory Pool: module to organise incoming transactions.
    \item Block Proposer: module to create block proposals for state keeper
    \item Committer: module to store pending and completed blocks into the database
    \item API: module to allow users to interect with zkSync Lite to query block data or submit transactions.
    \item Ethereum Sender: module to sync the operations on zkSync Lite with the Ethereum blockchain. It makes sure that the L1 transactions zkSync Lite created (such as committing a block on-chain) are executed on-chain in the correct order. 
\end{enumerate}

\subsubsection{Prover.}
The Prover's only job is to create block proofs given a block's transaction witnesses. It regularly polls the Server for blocks that do not have a corresponding SNARK. When a new block is available, Server sends the block's witnesses so the Prover can begin creating the block proof. Once finished, the Prover returns the SNARK to the Server and the server sends it to the on-chain Smart Contract to be verified.

\section{zkSync Lite Transaction Flow}
Below we describe the transaction flow on zkSync. First, we provide an end-to-end description from L2 transaction submission to L2 block finalisation on-chain. Then, we zoom in on the Server to describe the flow within the Server in details.

\subsection{Overall Transaction Flow}
\begin{figure}[ht]
\centering
\includegraphics[width=1\textwidth]{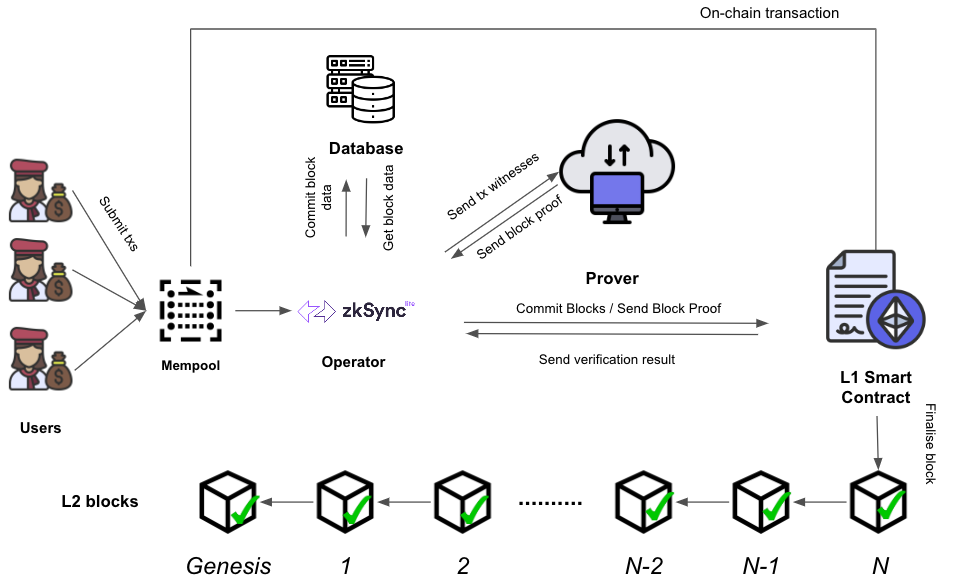}
\caption{An illustration of transaction flow within zkSync Lite from transaction submission to L2 block finalisation.}
\label{fig:zksync_lite_design}
\end{figure}

Referring to Figure \ref{fig:zksync_lite_design} for a simplified representation of zkSync Lite. When a user submits a transaction, it is placed into the memory pool (mempool) waiting to be collected by the Server. The server periodically collects a queue of transactions from the mempool, in submission order, and puts them into blocks. After the blocks are formed, they are committed to the L1 Smart Contract and stored in the database. At this moment, although the block information is on-chain, they are not finalised. These blocks in this state are known as the ``committed block".

At the same time, available Provers poll the Operator for proof generation jobs. When there are blocks without a proof, the Operator will generate and send the block witnesses to the Prover, who will use the witnesses to generate and return the block proof. Once the operator receives the block proof, it will send it to the L1 Smart Contract for verification. 

The Verifier contract verifies the block proof along with the committed block data. The L2 Smart Contracts updates the block's from committed to finalised when the proof is validated. 

For priority transactions (listed in Section \ref{appx:zksync_tx_types}) that are submitted directly to the L1 Smart Contract, they are tracked by the Operator and added to the mempool into a priority queue.

\subsection{Transaction Flow within Server}\label{appx:tx_flow_server}
\begin{figure}[ht]
\centering
\includegraphics[width=0.8\textwidth]{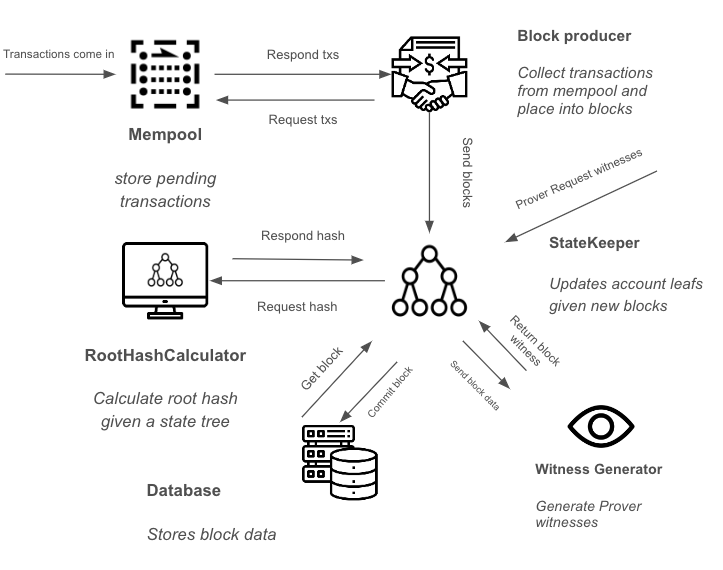}
\caption{An illustration of transaction flow zkSync Lite server.}
\label{fig:server_design}
\end{figure}

Looking specifically into the Server shown in Figure \ref{fig:server_design}, as blocks are created by the block producer, they are sent to the State Keeper. The State Keeper processes the transactions in the blocks and update the accounts' balances accordingly. Although it stores the Account Tree, it does not update the Account Tree's root hash. It delegates the computation intensive job to the Root Hash Calculator, where the re-hashing of the tree is done. Once a block is completed with a root hash, it is committed to the database. 

As the Prover polls for committed blocks, the Witness Generator will generate transaction witnesses and send to the Prover. 

\section{zkSync Transaction Types}\label{appx:zksync_tx_types}
As mentioned above, zkSync Lite supports a number of transaction types. Here, we give a brief description of these transaction types. Full descriptions can be found in~\cite{zkSync_lite}. There are two main categories of transactions on zkSync: normal and priority transactions. Priority transactions are handled by the operator differently during the L2 block creation process. Given a queue of transactions from the mempool, and an operator continually placing transactions into blocks, as soon as a priority transaction is processed, the current block is sealed and committed regardless of remaining gas in the block. 

\subsection{From Transactions to Leaf Operations}\label{appx:leaf_operations}
One or more accounts can be affected as a result of a transaction. For example, a transfer transaction adds to the receiver's balance, as well as deducting from the sender's balance. To make the account leaf updates atomic, zkSync breaks down each type of transactions into their a number of \textit{leaf operations}. Each operation only affects one account leaf at a time. 

In the following sections, as we describe the transaction types, we include the number of operations to which they can be broken down. 

\subsection{Normal Transaction Types}
\begin{enumerate}
    \item \texttt{Transfer}: Transfer funds between rollup accounts. It translates to two \texttt{SMT.UpdateLeaf} operations. The first decreases the sender balance, and the second increases the receiver balance.
    \item \texttt{Transfer to new}: Transfer funds to a new account. This transaction type is derived from \texttt{Transfer} and happens when the \textsf{to\_account} doesn't exist in the \texttt{AccountTree}. Before the transfer of funds, a new account will be created for \texttt{to\_account}. It translates to an \textit{update} and an \texttt{SMT.InsertLeaf} operation. The first one decreases sender balance, and the second one inserts a new account leaf. 

    \item \texttt{Withdraw}: Withdraw funds from the L2 account to the indicated Ethereum address. It translates to an \texttt{SMT.UpdateLeaf} operation where the balance of the withdrawal account is decreased.

    \item \texttt{Withdraw NFT}: Withdraw NFT from the L2 account to the indicated \\Ethereum address. It translates to two \texttt{SMT.UpdateLeaf} operations. The first removes the NFT from the owner account and the second removes the NFT from the creator's account. 

    \item \texttt{Mint NFT}: Mint an NFT token inside L2.  It translates to two \texttt{SMT.UpdateLeaf} operations. The first adds the NFT to the receiver's account, and the second updates the creator's account. 

    \item \texttt{Change pubkey}: Change the public key used to authorise transactions for an account. This can be useful when a user wishes to delegate the account to another user or Smart Contract wallet with a different Ethereum address without the need to expose their own private key. It translates to an \texttt{SMT.UpdateLeaf} operation on the sender's account where the public key is updated.  

    \item \texttt{Forced Exit}: Withdraw funds from L2 accounts without the signing key to the appropriate L1 address. These accounts are known as \textit{unowned} accounts. It translates to \texttt{SMT.UpdateLeaf} and \texttt{SMT.RemoveLeaf} operations. The first up decreases the sender's balance, and the second one removes the account leaf and replaces it with a default node. 

    \item \texttt{Swap}: Perform an atomic swap of ERC-20 tokens between two L2 accounts at a defined ratio. Its operations are similar to the \texttt{transfer} transaction type. 
\end{enumerate}

\subsection{Priority Transaction Types}
\begin{enumerate}
    \item \texttt{Deposit}: Deposit funds from Ethereum to L2. The funds are sent to the zkSync Lite Smart Contract, which informs the operator to include a deposit transaction in the next block. A new account is created if necessary. It may translate to an \texttt{SMT.UpdateLeaf} operation, or an \texttt{SMT.InsertLeaf} operations. The operation is \texttt{SMT.UpdateLeaf} when the account already exists. On the other hand, the operation is \texttt{SMT.InsertLeaf} when a new account needs to be created. 
    \item \texttt{Full exit}: In the event that a user thinks the operator has censored their transactions, they can submit a \texttt{Full exit} transaction directly to the Smart Contract. The operator will process the transaction accordingly. Its operations are the same as \texttt{Forced exit}
\end{enumerate}

In the event that a priority transaction has not been processed for more than a week, the system will enter the \textsf{exodus mode} and the operators will stop working, and every user can use an exit tool \footnote{\url{https://github.com/matter-labs/zksync/tree/master/infrastructure/exit-tool}} to withdraw their asset by submitting a proof of balance to the L1 smart contract. 

\section{zkSync Lite Sparse Merkle Tree Usage}
zkSync Lite uses the SMT in three separate places as a data accumulator. They are the account tree, the circuit account tree, and the balance tree. 

\subsubsection{Account Tree.}
The Account Tree\footnote{\url{https://github.com/matter-labs/zksync/blob/master/core/lib/types/src/lib.rs\#L84}} is a binary SMT of depth 24. It is the main data structure that stores the state of the zkSync Lite accounts. Its leaves are the accounts on zkSync. The leaf hash is the rescue hash~\cite{aly2020design} of an account's fields concatenated in their respective little-endian bit representation. 

The leaf indices are the same as the account IDs, which are mapped to account addresses. Empty leaves are replaced with a default hash. 

\subsubsection{Circuit Account Tree.}
The purpose of the Circuit AccountTree is to generate compatible transaction witnesses so that the Prover can create the block proof. The Circuit Account Tree is structured similar to the AccountTree except for two main differences: 1) account data are encoded as field elements and 2) each account uses an SMT to track balances for each type of token (Balance Tree) instead of using a simple hash map.
The Circuit AccountTree is derived from AccountTree. 

\subsubsection{Balance Tree.}
As mentioned above, the Balance Tree is a part of the account leaves in the Circuit Account Tree. It is an SMT of depth 8. Each leaf in the Balance Tree represents the balance of the token with the id the same as the leaf index. 

\end{document}